\documentclass[fleqn, 3p, twocolumn]{elsarticle}

\usepackage[
bookmarks=true, colorlinks, linkcolor=blue, urlcolor=cyan, citecolor=red]{hyperref}

\usepackage{cancel}
\usepackage{color}
\usepackage{graphicx}

\usepackage{latexsym}
\usepackage{amsfonts}
\usepackage{amssymb}
\usepackage{amsmath}
\usepackage{bm}
\usepackage{mathrsfs}
\usepackage{slashed}

\usepackage{dcolumn}

\newcommand{\nn}{\nonumber}

\newcommand{\sla}{ \slashed }

\newcommand{\ep}{ \epsilon }

\newcommand{\gam}{ \gamma }

\newcommand{\Lm}{ \Lambda }

\newcommand{\bq}{{\bm{q}}}

\newcommand{\1}{\mbox{1}\hspace{-0.25em}\mbox{l}}

\newcommand{\sgn}{ {\rm sgn} }
\newcommand{\prj}{ {\mathcal P} }
\newcommand{\Q}{ {\mathcal Q} }

\newcommand{\M}{ {\mathcal M} }

\newcommand{\dyn}{ {\rm dyn} }
\newcommand{\LLL}{ {\rm LLL} }

\newcommand{\para}{ \parallel}

\newcommand{\beq}{\begin{eqnarray}}
\newcommand{\eeq}{\end{eqnarray}}

\usepackage{comment}
\usepackage{epstopdf}

\def\simge{\mathrel{%
   \rlap{\raise 0.511ex \hbox{$>$}}{\lower 0.511ex \hbox{$\sim$}}}}
\def\simle{\mathrel{
   \rlap{\raise 0.511ex \hbox{$<$}}{\lower 0.511ex \hbox{$\sim$}}}}
\def\bigs{\mathrel{
   \rlap{\raise 0.531ex \hbox{$>$}}{\lower 0.531ex \hbox{$<$}}}}

\usepackage[normalem]{ulem}  

\date{}

\begin{document}
\begin{frontmatter}
\title{Anatomy of the magnetic catalysis by renormalization-group method}
\author[Fudan] {Koichi Hattori}
\author[KEK,Soken] {Kazunori Itakura}
\author[Yagami,Hiyoshi] {Sho Ozaki}
%

\address[Fudan] {Physics Department and Center for Particle Physics and Field Theory, Fudan University, Shanghai 200433, China}
\address[KEK] {KEK Theory Center, Institute of Particle and Nuclear Studies, High Energy Accelerator Research Organization, \\1-1, Oho, Ibaraki, 305-0801, Japan}
\address[Soken] {Graduate University for Advanced Studies (SOKENDAI), 1-1 Oho, Tsukuba, Ibaraki 305-0801, Japan}
\address[Yagami] {Department of Physics, Keio University, Kanagawa 223-8522, Japan}
\address[Hiyoshi] {Research and Education Center for Natural Science, Keio University, Kanagawa 223-8521, Japan}

\begin{abstract}
We first examine the scaling argument for a renormalization-group (RG) analysis 
applied to a system subject to the dimensional reduction in strong magnetic fields, 
and discuss the fact that a four-Fermi operator of the low-energy excitations is marginal 
irrespective of the strength of the coupling constant in underlying theories. 
We then construct a scale-dependent effective four-Fermi interaction as a result of screened photon exchanges at weak coupling, 
and establish the RG method appropriately including the screening effect, 
in which the RG evolution from ultraviolet to infrared scales 
is separated into two stages by the screening-mass scale. 
Based on a precise agreement between the dynamical mass gaps obtained 
from the solutions of the RG and Schwinger-Dyson equations, 
we discuss an equivalence between these two approaches. 
Focusing on QED and Nambu--Jona-Lasinio model, 
we clarify how the properties of the interactions manifest themselves in the mass gap, and point out an importance of respecting the intrinsic energy-scale dependences in underlying theories for the determination of the mass gap. 
These studies are expected to be useful for a diagnosis of the magnetic catalysis in QCD. 

\end{abstract}

\end{frontmatter}

\section{Introduction}

Strong magnetic fields confine charged fermions in the lowest Landau levels (LLLs), and they enjoy the properties of the (1+1)-dimensional chiral fermions with the dispersion relation $ [e{\bm B} = (0,0,eB), \ eB>0] $: 
\begin{eqnarray}
\ep^{R/L}_{\rm LLL} =  \pm p_z
\, .
\label{eq:e_LLL}
\end{eqnarray}
Intuitively, this is a consequence of the formation of the small cyclotron orbit with the radius $\sim 1/|eB|^{1/2} $ and the residual free motion along the field. It turned out that this {\it dimensional reduction} gives rise to rich physics phenomena. Especially, the {\it magnetic catalysis} of the chiral symmetry breaking and the {\it chiral magnetic effect} have been addressed by many authors  (see, e.g., Refs.~\cite{Shovkovy:2012zn, Miransky:2015ava}
and Refs.~\cite{Kharzeev:2013ffa, Kharzeev:2015znc, Huang:2015oca, Hattori:2016emy} for reviews). 

The clear statement on the physical mechanism of the magnetic catalysis was due to Gusynin, Miransky, and Shovkovy in terms of a simple four-Fermi interaction~\cite{Gusynin:1994xp}. 
By solving the gap equation of the NJL model, they found a mass gap 
\beq
m_{\rm{dyn}}
&=& \sqrt{eB} \, {\rm{exp}} \left( - \frac{\pi}{  \rho^{}_{\rm LLL} G_{\rm NJL}  } \right),
\label{DynamicalMassNJL}
\eeq
where $\rho^{}_{\rm LLL}$ and $G_{\rm NJL} $ are the density of states in the LLL 
and a dimensionful coupling constant of the four-Fermi interaction, respectively. 
Their core observation is seen in the similarity between 
the mass gap and the energy gap of superconductivity which is given by 
$\Delta \sim \omega^{}_{\rm{D}} \, {\rm{exp}} [ - c^{\prime} /( \rho^{}_{\rm{F}} G^\prime )]$ 
with $\omega^{}_{\rm{D}}$ and $\rho^{}_{\rm{F}}$ 
being the Debye frequency and the density of states near the Fermi surface, respectively. 
Also, $  G^\prime$ and $c^{\prime}$ are a coupling constant and a positive number, respectively. 
In fact, this similarity is originated from the dimensional reduction 
in the low-energy domains of the both theories, i.e., 
in the LLL and in the vicinity of the Fermi surface. 

We can clearly see the consequence of the dimensional reduction 
by focusing on QED in 
the weak coupling regime. 
From the rainbow approximation of the Schwinger-Dyson (SD) equation, the mass gap was obtained as 
\begin{eqnarray}
m_\dyn \simeq \sqrt{eB}\, \exp \left( - \frac{\pi}{2} \sqrt{\frac{\pi}{\alpha}} \right)
\label{eq:QED_unsc}
\, ,
\end{eqnarray}
with an unscreened photon propagator in the early studies~\cite{Gusynin:1995gt, Gusynin:1995nb, Leung:1996qy}, 
and also 
\begin{eqnarray}
m_\dyn \simeq \sqrt{2eB} \, \alpha^{1/3} \exp \left\{ - \frac{\pi}{\alpha  \log(C \pi /\alpha)} \right\}
\label{eq:QED_sc}
\, ,
\end{eqnarray}
with a screened photon propagator~\cite{Gusynin:1998zq, Gusynin:1999pq}. 
Here, $\alpha=e^2/4\pi$ and $C$ is a certain constant of order one. 
The constant was analytically obtained as $C=1$ 
when the momentum dependence of $ m_\dyn $ is neglected. 
The authors of Refs.~\cite{Gusynin:1998zq, Gusynin:1999pq} observed that {\it the gap equation always has a nontrivial solution irrespective of the size of the coupling constant}, indicating that the strong magnetic fields cause the dynamical symmetry breaking without support of any other nonperturbative dynamics. This reminds us of the well-known fact that any weak attractive interaction causes superconductivity. 

Our main assertion in this Letter is that all these aspects of the magnetic catalysis can be 
understood with the Wilsonian renormalization group (RG) analysis. 
We will show that the emergence of the dynamical mass gap is 
informed from the RG flow for the effective four-Fermi operator that goes into the Landau pole. 
Bearing it in mind that four-Fermi operators are irrelevant in ordinary (3+1)-dimensional systems, 
we will clearly see from the RG point of view that the magnetic catalysis of the dynamical symmetry breaking 
is intimately related to the dimensional reduction. 
Our approach shares the philosophy with the analysis of 
(color) superconductivity by the RG method \cite{Polchinski:1992ed, 
Evans:1998ek, Evans:1998nf, Schafer:1998na, Son:1998uk, Hong:1998tn, Hong:1999ru, Hsu:1999mp}. 

Our ultimate goal is to consistently understand the enhancement of the chiral symmetry breaking 
at zero or low temperature, and the inverse magnetic catalysis near 
the chiral phase transition temperature in QCD. 
There has been a discrepancy between the estimates of the chiral condensate 
from the lattice QCD simulation and typical model calculations~\cite{Bali:2012zg}. 
For the magnetic catalysis and the inverse catalysis 
to be compatible with each other, it appeared to be important to explain 
a mechanism which makes the dynamical mass gap stay as small as the QCD scale $ \Lambda_{\rm QCD} $ 
even in a strong magnetic field $ eB \gg \Lambda_{\rm QCD}^2 $ \cite{Kojo:2012js} (see also Refs.~\cite{Watson:2013ghq, Mueller:2014tea, Hattori:2015aki}). 
The method of renormalization group is a potentially useful tool 
to obtain a clear insight on this issue on the basis of 
an argument of the hierarchy which we will elaborate in the present paper. 

However, to the best of our knowledge, 
even the correct form of the mass gap in weak-coupling gauge theories 
has not been obtained by the RG analyses in the presence of the screening effect.
Therefore, before discussing the strong-coupling regime in QCD, 
one should understand how the screening effects are reflected in 
the parametric form of the mass gap in a clear way. 
Moreover, it is a generic issue to establish a systematic way of 
including the screening effects in the RG analyses, 
which will be important in a variety of systems. 
Note, for example, that there was an issue of the color magnetic screening in the RG analysis on the color superconductivity \cite{ Son:1998uk}.

We will show that all of the results in Eqs.~(\ref{DynamicalMassNJL}), (\ref{eq:QED_unsc}), and (\ref{eq:QED_sc}) 
from the SD equations are precisely obtained from the solutions of the RG equations. 
Furthermore, we will clarify the origins of the overall factor of $  \sqrt{eB}$ and 
the exponents 
in the language of the RG method. 
We will find that the properties of the interactions in the model/theory 
are directly reflected in the parametric dependences of the dynamical mass 
on the coupling constant and the magnitude of $ eB $. 
Ultimately, these studies will be useful for a diagnosis 
of the magnetic catalysis in QCD. 
We will come back to this point with a brief comment on the perspective in the last section. 


More specifically, we will closely look into the screening effect on the photon propagator. 
It would be instructive to mention a successful application of the RG method 
to color superconductivity in dense quark matter, 
where an appropriate treatment of  the dynamical screening effect on the magnetic gluons 
was important for obtaining the correct magnitude of the gap~\cite{Son:1998uk, Hsu:1999mp}. 
We should also mention that the RG analysis of the magnetic catalysis at weak coupling 
was performed in Refs.~\cite{Hong:1996pv,Hong:1997uw}. 
Also, the magnetic catalysis in QCD was investigated 
on the basis of both the SD and RG equations in Ref.~\cite{Mueller:2015fka}. 
However, roles of the screening effect arising from the quark loop 
in the magnetic field have not been identified thus far, 
and we are not aware of the RG analysis in the literature 
of which the result agrees with that from the SD equation (\ref{eq:QED_sc}).

As we will discuss later in more detail, the screening effect should be appropriately incorporated 
in the derivation of the RG equation, since the screening mass sets an intrinsic energy scale 
of the underlying theory in between the ultraviolet and infrared regimes. 
The essential technique was recently developed for the analysis of the RG flow 
in ``magnetically induced QCD Kondo effect"~\cite{Ozaki:2015sya}. 
In the present Letter, we will show that the same technique successfully works 
for the analysis of the magnetic catalysis at weak coupling.


The structure of this Letter is the following. We first show the connection between the magnetic catalysis and the dimensional reduction which can be understood from a simple discussion 
of the scaling dimensions. 
Next, we construct an effective four-Fermi interaction from the underlying weak-coupling theory, i.e., QED, and appropriately include the energy-scale dependence of the tree-level interaction. 
Based on these discussions, 
we derive the RG equations and obtain the dynamical mass gap from their solutions. 
We confirm that the energy-scale dependence of the interaction 
is necessary for obtaining the correct form of the gap, 
which was however missing in the previous analyses. 
Finally, we discuss the correspondences between the RG and SD analyses, 
and the crucial roles of the photon/gluon propagators in the magnetic catalysis. The derivation of the RG equation 
is briefly summarized in an appendix.

\section{Infrared scaling dimensions}

\label{sec:dense-mag}

We begin with looking into an analogy between the systems in the strong magnetic field and at high density. In the presence of a large Fermi sphere, the low-energy excitations near the Fermi surface 
show the dimensional reduction: The two-dimensional phase space tangential to the large Fermi sphere is degenerated, and the energy dispersion depends only on the momentum normal to the sphere. Then, the dimensional reduction enhances the infrared (IR) dynamics, 
leading to the instabilities near the Fermi surface. Based on the analogy with
this mechanism, Gusynin et al. clearly pointed out that the chiral symmetry breaking occurs in the strong magnetic field no matter how weak the coupling is \cite{Gusynin:1994xp}. 

One can see possible emergence of the IR instability from a simple argument of the scaling dimensions. The kinetic term for the LLL reads 
\begin{eqnarray}
&& \hspace{-1.2cm}
S_{\rm LLL}^{\rm kin} 
\label{eq:L_LLL}
\\
&&\hspace{-0.8cm}=
\int \!\! dt \!\! \int \!\! dp_z \bar \psi^{}_{\rm LLL} (p_z) (i\partial_t \gam^0 - p_z \gam^3 ) \psi^{}_{\rm LLL} (p_z)
\nn
,
\end{eqnarray}
where we have suppressed the label specifying the location of the cyclotron center on the transverse plane. From this kinetic term, one can find the IR scaling dimension of the LLL fermion field when the excitation energy goes down toward zero 
as $ \ep^{}_{\rm LLL} \to s \ep^{}_{\rm LLL} $ ($t \to s^{-1}t$) with $s<1$. 
Since the LLL fermion has the (1+1) dimensional dispersion relation (\ref{eq:e_LLL}), 
the longitudinal momentum $  p_z$ also scales as $ p_z \to s p_z $. 
On the other hand, the transverse momentum does not scale, because it serves just as the label of the degenerated states and does not appear in the dispersion relation (\ref{eq:e_LLL}). Therefore, when the kinetic term (\ref{eq:L_LLL}) is invariant under the scale transformation, the LLL fermion field scales as $ s^{-1/2} $ in the low-energy dynamics. 

Bearing this in mind, we proceed to the effective four-Fermi operator in the LLL: 
\beq
&& \hspace{-0.9cm}
{S}_{\rm LLL}^{\rm{int}} = \!\!
\int\!\! dt\!\! \prod_{i=1,2,3,4} \int \!\! dp_{z}^{(i)}\, G\, \delta( p_{z}^{(1)} + p_{z}^{(2)} - p_{z}^{(3)} - p_{z}^{(4)} )
\nn
\\
&& \hspace{-0.7cm} \times 
\left[\bar{\psi}_{\rm{LLL}}(p^{{(2)}}_{z}) \gamma^{\mu}_{\parallel} \psi_{\rm{LLL}}(p^{{(4)}}_{z})\right] \!\!
\left[ \bar{\psi}_{\rm{LLL}}(p_{z}^{(3)}) \gamma_{\parallel \mu} \psi_{\rm{LLL}} (p^{(1)}_{z}) \right]
\!\! ,
\nn
\\
\label{four_int_B} 
\eeq
where $G$ is an effective coupling constant and $ \gam_\para^\mu =(\gam^0,0,0,\gam^3) $. 
The transverse momenta are again not written explicitly. Note that the delta function has the scaling dimension $s^{-1}$ due to the dimensional reduction in Eq.~(\ref{eq:e_LLL}). Thus, we find that the four-Fermi interaction term has the scaling dimension $s^{0}$, meaning that the four-Fermi operator is {\it marginal} in the strong magnetic field \cite{Hong:1996pv, Hong:1997uw}.

This result suggests that the chiral symmetry could be broken 
even in weak-coupling theories like QED, 
which is just like the well-known fact of superconductivity that 
any weak attraction induces the BCS instability. 
Essentially, the magnetic catalysis occurs only for the dimensional reason. 
Below, we will explicitly see the indication of the magnetic catalysis by means of the RG approach. 
The logarithmic quantum correction will drive the effective coupling constant $  G$ into the Landau pole.

\section{Effective interactions from underlying theories}

\begin{figure}
\vspace{-1.2cm}
     \begin{center}
              \includegraphics[width=0.8\hsize]{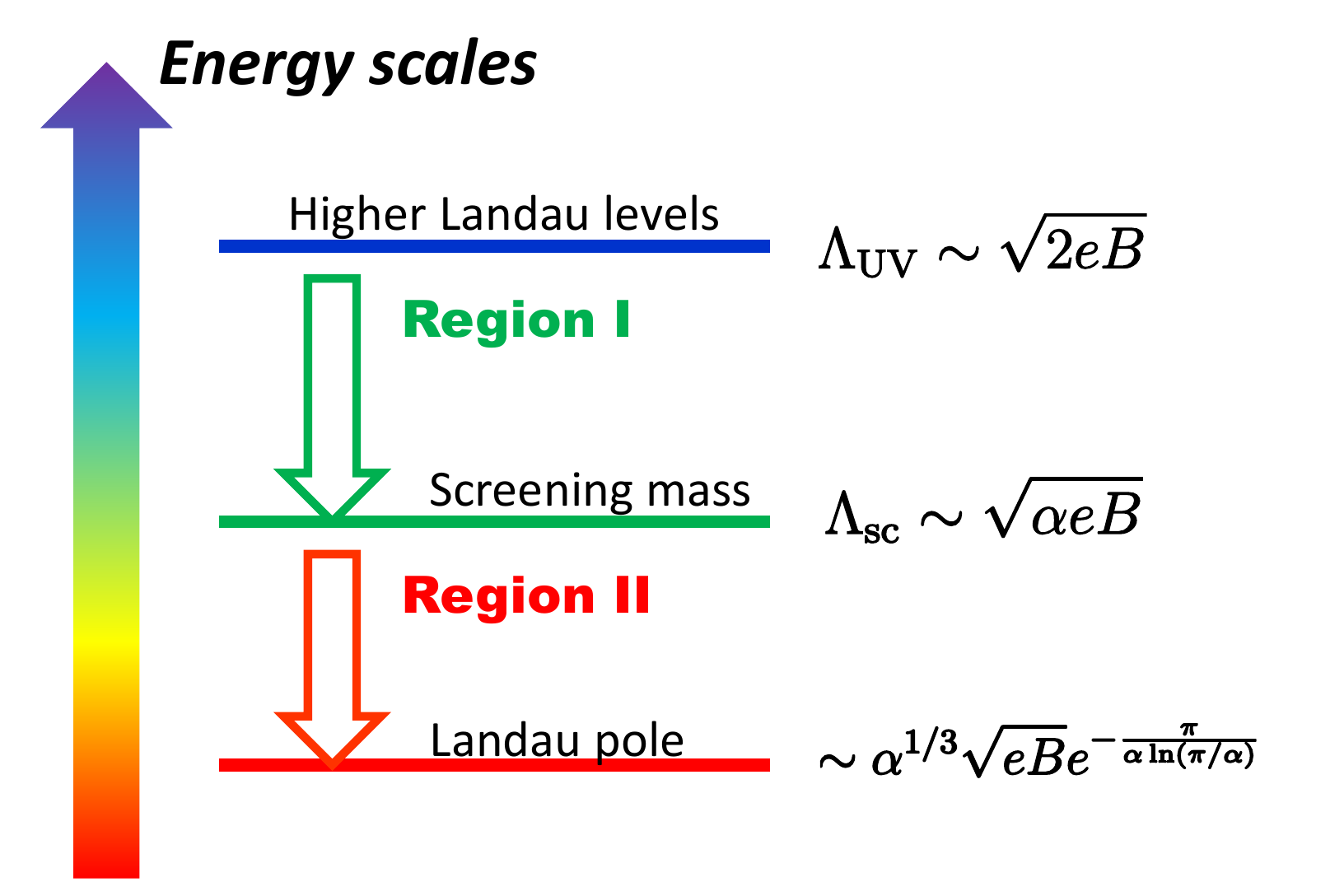}
     \end{center}
\vspace{-0.6cm}
\caption{Energy scales appearing in the RG evolution. 
The relevant region from the UV cutoff down to the Landau pole 
is divided into two regions by the scale of the screening mass.}
\label{fig:RG-MC}
\end{figure}

To build a bridge between the underlying gauge theories and the scaling-dimension argument for the four-Fermi operator, we first construct an effective four-Fermi interaction based on the underlying theory. 
It is very crucial to correctly take into account the energy-scale dependence of the interactions in gauge theories which are mediated by gauge bosons. Elaborating this point, we will obtain the correct form of the dynamical mass, and will show that the reliable effective theory gives a different RG evolution from 
that in the scale-independent four-Fermi interaction model 
which do not respect the scale dependences in the underlying theory.

In QED as the underlying theory, we shall introduce the photon propagator in the strong magnetic field that has a screening mass \cite{Gusynin:1998zq, Gusynin:1999pq, Fukushima:2011nu, 
Hattori:2012je, Fukushima:2015wck, Hattori:2017xoo}. 
The explicit form in a non-covariant gauge reads~\cite{Gusynin:1998zq, Gusynin:1999pq} 
\beq
i \mathcal{D} _{\mu \nu} (q)
&=& \frac{ g_{\parallel}^{\mu \nu} }{ q^{2} - m_\gam^2  }
\label{photon_prop_inp}
\\
&&
+ \frac{ g_{\perp}^{\mu \nu} }{ q^{2} }  
- \frac{ q_{\perp}^{\mu} q_{\perp}^{\nu} 
+ q_{\perp}^{\mu} q_{\parallel}^{\nu} + q_{\parallel}^{\mu} q_{\perp}^{\nu} }{ (q^{2})^{2} }
\nn
\, ,
\eeq
where $ g_\para^{\mu\nu}= {\rm diag} (1,0,0,-1) $, $ g_\perp^{\mu\nu} = {\rm diag} (0,-1,-1,0) $, 
and $q_{\para,\perp}^\mu = g_{\para,\perp}^{\mu\nu} q_\nu  $. 
The screening mass $m_{\gamma}^{2} = 2 \alpha eB / \pi$ 
comes from the one-loop photon self-energy composed of the LLL fermion lines 
in the vanishing frequency limit.\footnote{ 
While we discuss the one-flavor case, 
the extension to the multi-flavor cases just results in the overall summation of the charge in the screening mass. }
The other terms are coupled to neither the LLL fermion loop nor the scattering LLL fermions 
because the LLL fermion current, $ j^\mu_\LLL = \bar \psi^{}_\LLL \gam^\mu_\para \psi^{}_\LLL$, 
is longitudinal to the magnetic field. 
Thus, those terms are completely irrelevant in the present discussion.

Now, we note that the dispersion relation of the LLL fermion is given by 
the (1+1)-dimensional form (\ref{eq:e_LLL}), while the photons live in the ordinary four dimensions. 
Therefore, we define the effective coupling constant $ G $ by integrating out 
the transverse momentum in the photon propagator as \cite{Kojo:2012js, Hattori:2015aki, Ozaki:2015sya} 
\beq
G (q_\parallel^2) &\equiv& \int \frac{ d^2 \bq_{\perp}} { (2\pi)^{2} }
\frac{ (-ie)^{2} }{ q_{\parallel}^{2} - \bq_{\perp}^{2} 
- m_{\gamma}^{2} }\, {\rm{e}}^{ - \frac{ \bq_{\perp}^{2} }{ 2 eB } }
\label{effecG_magMC}
\, ,
\eeq
where the exponential factor comes from the transverse part of the fermion wave functions. 
This effective coupling allows us to identify the dimensionally reduced effective interaction discussed 
in Eq.~(\ref{four_int_B}) which however possesses an appropriate  
energy dependence as we will see below. 

\begin{figure}
\vspace{-1.2cm}
     \begin{center}
              \includegraphics[width=1\hsize]{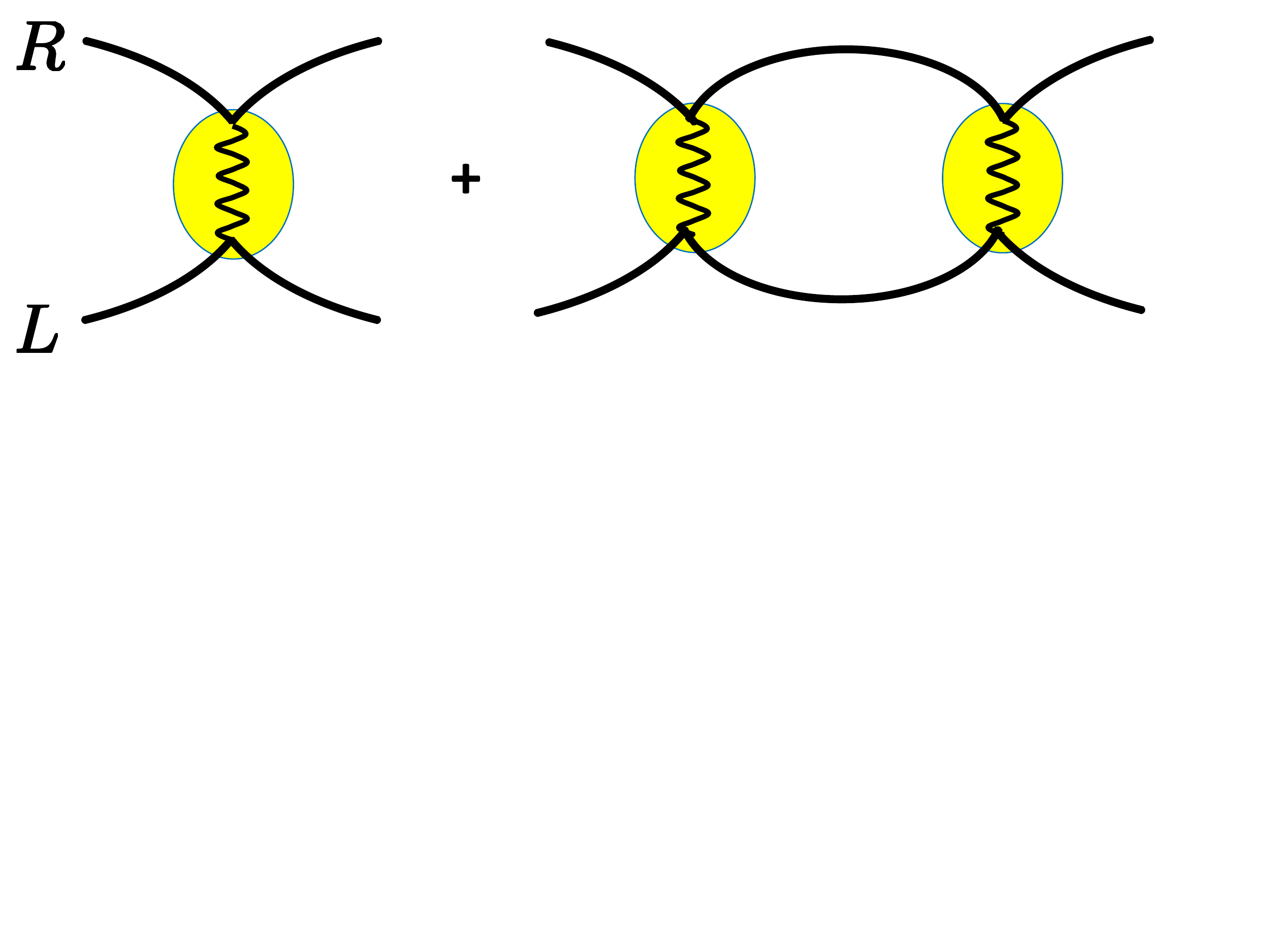}
     \end{center}
\vspace{-0.7cm}
\caption{Scattering diagrams contributing to the RG flow in the magnetic catalysis. 
Yellow blobs capture the the running effective coupling constants which 
are defined in terms of the exchanged gauge-boson propagators in the weak-coupling theory. }
\label{fig:MC-diagrams}
\end{figure}

Importantly, we have an intrinsic energy scale given by the screening mass $ \Lm_{\rm sc} \equiv m_\gam $ 
which cuts off the IR region of the transverse momentum integral. 
Therefore, the result of the integral depends on 
if the scale of interest $  \Lm$ is larger than $ \Lm^{}_{\rm sc} $ [Region (I): $ \Lm > \Lm^{}_{\rm sc} $]
or is in the deeper IR region $ \Lm < \Lm^{}_{\rm sc} $ [Region (II): $ \Lm < \Lm^{}_{\rm sc} $]. 
As summarized in Fig.~\ref{fig:RG-MC}, one needs to examine the RG evolution in these regions separately. 
On the other hand, the upper boundary of the integral is, in the both cases, given by 
$ \Lm^{}_{\rm UV} \equiv \sqrt{2eB} $ which appears in the exponential factor 
of Eq.~(\ref{effecG_magMC}) 
and also corresponds to the energy scale where the higher Landau levels start to contribute. 
Based on this scale-dependent four-Fermi interaction, 
we investigate the RG evolution in the next section. 


\section{RG analysis at weak coupling}

\label{sec:RG_MC}

We derive the RG equation for the coupling constant of the effective four-Fermi interaction 
in terms of the Wilsonian renormalization group. 
More specifically, we compute the scattering amplitudes for the fermion and antifermion pair 
that forms the chiral condensate (see Fig.~\ref{fig:MC-diagrams}), and integrate out 
the excited states. 
As we have already learned that the four-Fermi operator is marginal in the LLL, 
we anticipate that the loop integral in the scattering amplitude generates a logarithm, 
and renormalizing the effective coupling constant with this logarithm will 
drive the system toward a strong coupling regime. 

As shown in Fig.~\ref{fig:MC-diagrams}, the leading-order (tree-level) scattering amplitude 
is given by the one-photon exchange diagram: 
\beq
\mathcal{M}_0 = G (q_\para^2)
\label{tree_amp_magneticfields}
\, .
\eeq
The relevant scattering channels for the formation of the chiral condensate 
are those between the pairs carrying the opposite chiralities, 
so that the two relevant spinor structures are given by 
$[ \, \bar{u}_{\rm{R/L}} (p^{(3)}_{\parallel})  \gamma_{\parallel \mu}u_{\rm{R/L}} (p^{(1)}_{\parallel}) \, ]\, 
[ \, v_{\rm{L/R}} (p^{(4)}_{\parallel} ) \gamma^{\mu}_{\parallel}  \bar{v}_{\rm{L/R}} (p^{(2)}_{ \parallel})  \, ]  
$ 
with the spinors of the fermion $ u $ and the antifermion $ v $ in the LLL. 
The scattering amplitudes are the same for the both channels, 
and below these trivial spinor structures will be suppressed for notational simplicity. 

Since the coupling constant $G(q_\para^2)$ has the scale dependence discussed in the previous section, 
the tree-level amplitude also contributes to the RG evolution. 
This situation is somewhat, though not exactly, similar to that in the color superconductivity 
in dense quark matter where the scale dependence of the dynamical screening mass 
in the tree-level diagram modifies the exponent in the critical temperature \cite{Son:1998uk}.

In the present case, when the energy and momentum scales of the fermion are of the order of $ \Lm $, 
the momentum transfer is $ - \Lm^2 \lesssim q_\para^2 \lesssim 0 $, 
where we are only interested in the space-like region contributing to the fermion scatterings. 
Now, when integrating out the fermionic degrees of freedom in the shell $ (\Lm-\delta\Lm ) \sim \Lm  $, 
we obtain the increment of $ \M_0(\Lm) $ as 
\begin{eqnarray}
\M_0(\Lm - \delta \Lm) - \M_0(\Lm ) 
= G( - (\Lm - \delta \Lm )^2 ) - G( - \Lm^2 ) 
\, .
\end{eqnarray}
In Region II, the dependence on $  \Lm$ goes away, 
because the screening scale $  \Lm_{\rm sc}$ dominates over $ \Lm $. 
However, in Region I, the screening mass is negligible, 
so that there is the dependence on $ \Lm $. 
Performing the transverse-momentum integral in Eq.~(\ref{effecG_magMC}), we have 
\begin{eqnarray}
G( - (\Lm - \delta \Lm )^2 ) - G( - \Lm^2 ) \sim 
\alpha \int ^{\Lm^2}_{(\Lm-\delta\Lm)^2} \frac{dq^2_\perp}{q^2_\perp}
\, .
\end{eqnarray}
Therefore, the result at the tree level is summarized as 
\beq
&&
\M_0(\Lm - \delta \Lm) - \M_0(\Lm ) 
  \nn
  \\
 && \hspace{.5cm}
\simeq \left\{  \begin{array}{ll}
2 {\alpha} \, {\rm{log}} \left( \frac{ \Lambda }{ \Lambda - \delta \Lm } \right) & 
{\rm Region \ I}
\\
0 &  
 {\rm Region \ II}
  \end{array} \right. .
  \label{effeG_MC2}
\eeq
This intrinsic scale dependence of the effective coupling constant 
partly drives the RG evolution, which should be taken into account in the RG equation below.

\begin{figure}[t]
	\vspace{-1.cm}
	\begin{center}
		\includegraphics[width=0.75\hsize]{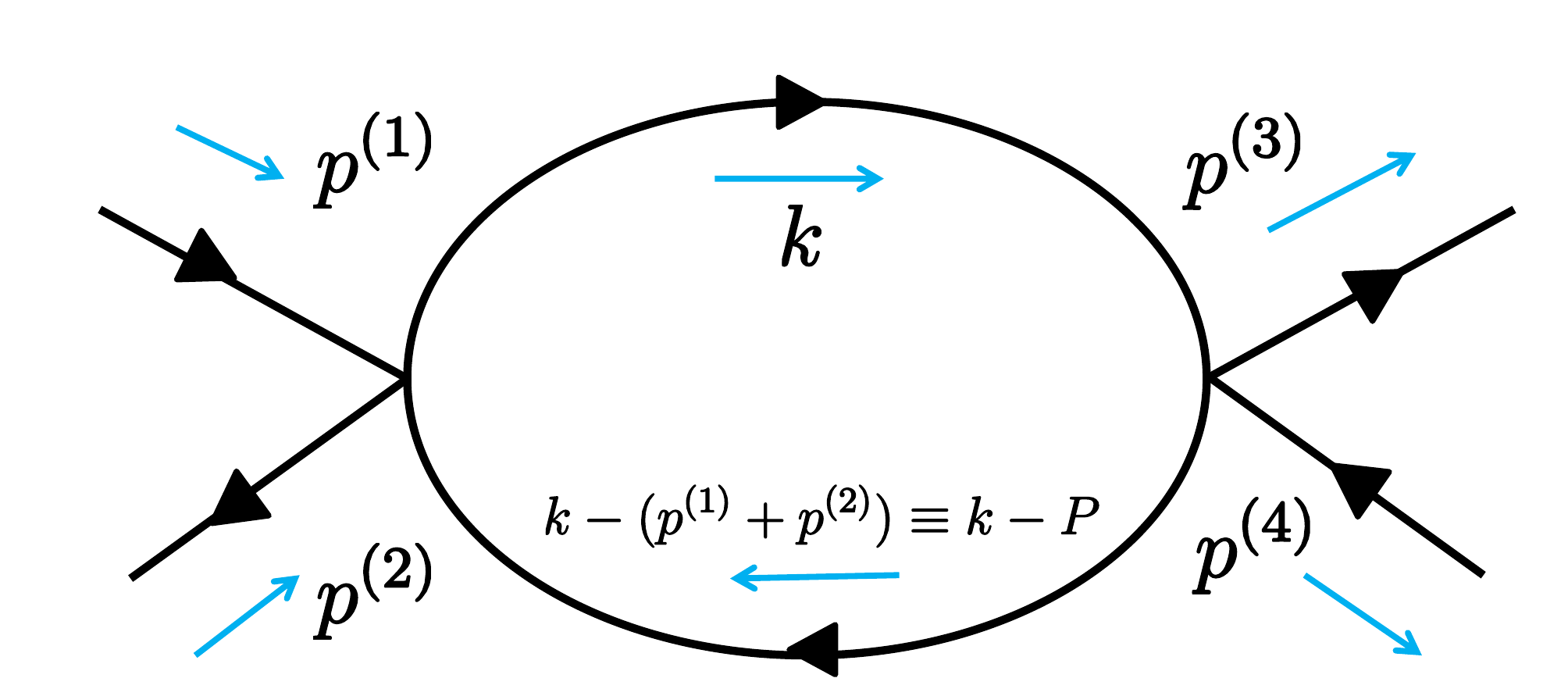}
	\end{center}
	\vspace{-0.5cm}
\caption{Diagram for the scattering between the particle and antiparticle.}
\label{fig:box}
\end{figure}

Next, let us proceed to the one-loop amplitude shown in Figs.~\ref{fig:MC-diagrams} and \ref{fig:box}. 
It is written down as 
\beq
\label{Amp_1-loop}
&&
\hspace{-1cm}
-i \mathcal{M}_1
=  ( -iG )^{2} \int \frac{ d^{2} k_{\parallel} }{ (2\pi)^{2} }
\left[ \bar{u} (p^{(3)}_{\parallel}) \gamma_{\parallel}^ {\mu}
S ( k_{\parallel}  ) 
\gamma_{\parallel}^ {\nu} u(p^{(1)}_{ \parallel}) 
\right]
\nonumber 
\\
&& \hspace{1.8cm}\times 
\left[ v(p^{(4)}_{ \parallel})  \gamma_{\parallel \mu} 
S (k_{\parallel} - P_\parallel) 
\gamma_{\parallel \nu} \bar{v} (p^{(2)}_{ \parallel})\right] 
\, ,
\nn
\\
\eeq
where $ P_\para = p^{(1)}_{\parallel} + p^{(2)}_{ \parallel}  $. 
The LLL propagator in the Ritus basis is given by 
\begin{eqnarray}
S (k) =  \frac{i \sla k_\para}{k_\para^2 + i \epsilon} \prj_+
\label{eq:LLL-prop}
\, ,
\end{eqnarray}
with the spin projection operator $  \prj_\pm = ( 1\pm i  \gam^1 \gam^2)/2$ 
in the direction of the magnetic field. 

We are left with the two-dimensional loop integral in Eq.~(\ref{Amp_1-loop}), 
which is a natural consequence of the dimensional reduction. 
After performing the elementary $k^{0}$-integral as explained in 
\ref{sec:amp}, one finds the origin of the magnetic catalysis: 
\beq
\label{eq:kz}
\mathcal{M}_1= 
G^{2}\int \frac{ d k_{z} }{ 2\pi } \frac{ 1 }{ | k_{z} | } 
\, .
\eeq
The same result is obtained for the both scattering channels, so that 
we have again suppressed the trivial spinor structures which are common to the tree-level amplitude. 
The remaining one-dimensional integral has the anticipated logarithmic IR divergence. 
When the scale goes down from $ \Lm $ to $ \Lm - \delta \Lm $, 
the increment of the one-loop amplitude is found to be 
\beq
\mathcal{M}_1(\Lm -\delta \Lm)  - \mathcal{M}_1(\Lm) 
=
\frac{ G^{2}  (\Lm)  }{\pi}\log \frac{\Lambda }{\Lambda - \delta \Lm} 
\label{eq:MC-one-loop}
\, .
\eeq
The logarithm is absorbed by the renormalization of the effective coupling constant $  G$, 
which leads to the RG evolution.

From the scattering amplitudes in Eqs.~(\ref{effeG_MC2}) and (\ref{eq:MC-one-loop}), 
the RG equation is obtained as 
\begin{subequations}
\beq
&&\hspace{-1cm} 
\Lambda \frac{d}{d\Lambda}G(\Lambda)=-{2\alpha } -\frac{1}{\pi}  G^2(\Lambda)
\ \ \ \ \  
{\rm Region \ I}
\, , 
\label{RGeq-I_MC}
\\
&&\hspace{-1cm} 
\Lambda \frac{d}{d\Lambda}G(\Lambda)=-\frac{1}{\pi} G^2(\Lambda)
\hspace{1.4cm}  
{\rm Region \ II}
\label{RGeq-II_MC}
\, .
\eeq
\end{subequations}
The first term in Eq.~(\ref{RGeq-I_MC}) comes from the tree-level amplitude 
according to the intrinsic scale dependence of the coupling constant shown in Eq.~(\ref{effeG_MC2}). 
This term is absent in Eq.~(\ref{RGeq-II_MC}), 
because the soft momentum transfer is cut off by the screening mass. 
The clear distinction between these two regions will result in an important consequence, 
as first discussed in the analysis of magnetically induced QCD Kondo effect~\cite{Ozaki:2015sya}. 
The other terms in these equations, which come from the logarithmic contribution in Eq.~(\ref{eq:MC-one-loop}), have minus signs, so that these terms drive the RG flow toward the Landau pole 
when the interaction is attraction ($ G>0$) as explicitly seen below. 

When the scale is reduced from $\Lm_0$ to $\Lm$ ($ \Lm > \Lm^{}_{\rm sc}$) where the initial scale $\Lambda_0$ is of the order of $\sqrt{eB}$, we use the RG equation (\ref{RGeq-I_MC}) in Region I. 
With the initial coupling $G (\Lm_0) = \alpha \, {\rm{log}} (2 eB / \Lm_{0}^{2}) \ll1$ 
from Eq.~(\ref{effecG_magMC}), we find the solution as 
\beq
G(\Lambda)
\simeq \sqrt{ 2 \alpha \pi } \,\, {\rm{tan}} 
\left( - \sqrt{ \frac{  \alpha }{ 2 \pi } }\, {\rm{log}} \frac{ \Lambda^2 }{ 2eB}  \right)
\label{GI_MC}
\, .
\eeq
Here, the $\Lambda_0$ dependence appears in the higher-order 
terms of $\alpha$, and can be neglected. 
Then, we find the running coupling constant at the lower boundary of Region I as 
\beq
G(\Lm^{}_{\rm sc}) \simeq   
\alpha \, \log \frac{ 2eB }{ m_{\gamma}^{2} }
+ \frac{ \alpha^2 }{ 6\pi} \left( \log \frac{ 2eB }{ m_{\gamma}^{2} }\right)^3
\, ,
\label{solution_mgamma}
\eeq 
where we performed an expansion with respect to 
$\alpha \, {\rm{log}}( 2eB / m_{\gamma}^2) = \alpha \, {\rm{log}}( \pi / \alpha) \ll 1$. 
The leading term of order $\alpha  $ corresponds to 
the tree-level coupling constant (\ref{effecG_magMC}), while the subsequent term 
explains the growth of the coupling constant driven 
by the quantum effect in the scale region $\Lambda^{}_{\rm sc} < \Lambda < \Lambda_0$.


When the scale $\Lambda$ enters Region II, 
we solve the RG equation (\ref{RGeq-II_MC}) with the initial condition at $\Lambda_0 = \Lm^{}_{\rm sc}$ 
which was obtained from the RG evolution in Region I in Eq.~(\ref{solution_mgamma}). 
In this way, the evolutions in the two regions are smoothly connected. 
We find the solution in Region II as 
\beq
G( \Lambda) = \frac{ G (\Lm^{}_{\rm sc}) }{ 1 + \pi ^{-1} G(\Lm^{}_{\rm sc})  
\log ( \Lambda/ \Lm_{\rm sc}  )} 
\, .
\label{Solution_of_RegionII}
\eeq
Clearly, this solution has a Landau pole at 
\begin{eqnarray}
\Lm_{\rm IR} = 
\Lm_{\rm sc} \, {\rm e}^{-{\pi}/{G(\Lm_{\rm sc})} }
\label{IRScale_MC}
\, ,
\end{eqnarray}
which indicates the emergence of the dynamical IR scale. 
The presence of the strong four-Fermi interaction induces a minimum 
of the effective potential at a nonzero value of the chiral condensate. 
The associated dynamical mass is of the order of the emergent scale $ \Lm_{\rm IR}  $, 
while the order of the chiral condensate is given 
in combination with the transverse degeneracy factor $ \sim eB  \Lm_{\rm IR} $: 
Intuitively, 
the condensate is squeezed along the magnetic field within 
the size of the cyclotron motion $\sim 1/\sqrt{eB}  $, 
and has a number of copies distributed with the density $  eB/(2\pi) $ 
in the transverse plane~\cite{Kojo:2012js} (see also a discussion in the last section). 
Therefore, the scale of the dynamical mass gap is explicitly 
\beq
m_{\rm{dyn}}
&\simeq& m_{\gamma}\, {\rm{exp}} \left\{ - \frac{ \pi }{ \alpha \, {\rm{log}} (\pi / \alpha ) } + {\rm{log}} \left( \frac{ \pi }{ \alpha } \right)^{1/6} \right\} 
\nonumber \\
&=& \sqrt{ 2 eB } \, \alpha^{1/3} {\rm{exp}} \left\{ - \frac{ \pi }{ \alpha \, {\rm{log}}(\pi / \alpha ) } \right\}
\label{DynamicalMass}
\, ,
\eeq
where the first and second terms in the exponential correspond to those in Eq.~(\ref{solution_mgamma}).

Notice that, no matter how small the coupling constant in the underlying theory is, 
the solution in Eq.~(\ref{Solution_of_RegionII}) has the Landau pole 
as long as the interaction is attractive.  
This is consistent with the aforementioned observation 
made in the study of the SD equations \cite{Gusynin:1998zq, Gusynin:1999pq}. 
By the use of RG analyses, 
this fact is even more clearly seen in the RG flow informed by the beta function. 
In short, when the magnitude of a magnetic field increases, 
the effect of the magnetic field shifts the UV fixed point 
that determines the critical coupling strength for the chiral symmetry breaking. 
Eventually, the fixed point merges with the one at the origin, 
and the beta function is completely pushed out from the positive region, 
leaving a vanishing critical coupling strength and the beta function entirely in the negative region 
(see Refs.~\cite{Fukushima:2012xw, Scherer:2012nn} for pedagogical discussions). 
This occurs only for the dimensional reason, as also discussed 
in a little bit different way in Sec.~\ref{sec:dense-mag} and implied by the integral in Eq.~(\ref{eq:kz}), 
and thus is a generic consequence of the dimensional reduction. 
We can immediately read off the beta function from Eq.~(\ref{RGeq-I_MC}) and (\ref{RGeq-II_MC}) as 
\begin{subequations}
\begin{eqnarray}
\beta_{\rm I} (\Lambda) &=& - 2\alpha  - \frac{1}{\pi} G^2 (\Lambda)
\, ,
\\
\beta_{\rm II}  (\Lambda) & =&  - \frac{1}{\pi} G^2  (\Lambda)
\, ,
\end{eqnarray}
\end{subequations}
for Region I and II, respectively. 
Clearly, the beta function is negative for any value of the coupling strength $  G(\Lambda)$, 
indicating that the broken phase is favored at zero temperature 
regardless of the value of the coupling strength $ \alpha $.

We remark on the contributions of the higher Landau levels. 
As is clear from Eq.~(\ref{DynamicalMass}), the location of the Landau pole 
is exponentially smaller than the Landau level spacing $ \sim \sqrt{eB} $. 
Therefore, the higher Landau levels are decoupled from the low-energy dynamics of the LLL 
and should not have any significant impact on the dynamical mass, 
consistently to the previous observations \cite{Kojo:2013uua}. 
Also, for sufficiently strong magnetic fields such that $ eB \gg m_f^2 $ 
with $  m_f$ being the intrinsic fermion mass, 
the dynamical mass, or $ \Lm_{\rm IR} $, emerges much earlier than the $ m_f $ 
when the energy scale goes down. 
Therefore, the $  m_f$ also should not have any significant impact.

\section{Discussions and concluding remarks}

Based on the RG analysis in the previous sections, 
we discuss an equivalence between the analyses by the RG and SD equations, 
and digest our RG analysis to identify the origins of the parametric forms of 
the dynamical mass gap (\ref{DynamicalMass}) in the screened QED 
as well as those in the NJL model and the unscreened QED [cf., Eqs.~(\ref{DynamicalMassNJL}) and (\ref{eq:QED_unsc})].

First, we would like to highlight the fact that 
our result (\ref{DynamicalMass}) has the same parametric form 
as that of the mass gap (\ref{eq:QED_sc}) obtained from the SD equation. 
This is not an accidental coincidence. 
The RG equations (\ref{RGeq-I_MC}) and (\ref{RGeq-II_MC}) correspond to 
the resummation of the ladder diagrams for the multiple photon exchanges. 
On the other hand, the analysis by the SD equation is based on 
the rainbow approximation, 
which holds only the planar diagrams of the fermion self-energy.
Cutting the intermediate fermion propagator in the SD equation provides 
the same ladder diagrams which we have included in the RG approach. 
Thus, solving the SD equation in the rainbow approximation is essentially equivalent to 
solving the RG equation in the leading order. 
Note also that the numerical constant $C$ in the exponent takes one in the both results. 
This agreement originates from a correspondence between the regularizations involved in the analyses. 
The result from the SD equation is obtained assuming that 
the fermion self-energy is a constant without a momentum dependence, that is, $ m_{\rm dyn} $, 
which regularizes an IR divergence of the SD equation in the (1+1) dimensions. 
This regularization corresponds to the sharp cutoff scheme in our RG analysis. 
The value of $C$ could depend on such a regularization 
as implied by a slightly different number $ C = 1.82 $ 
from the numerical result \cite{Gusynin:1998zq, Gusynin:1999pq}.


Now, we can identify the origins of the overall factors of $ \sqrt{eB} $, $ \alpha^{1/3} $, 
and the exponent in Eq.~(\ref{eq:QED_sc}) in the language of the RG method. 
We should first note that the dimensionful quantities enter 
in the RG evolutions (\ref{RGeq-I_MC}) and (\ref{RGeq-II_MC}) 
only through the energy scales $ \Lm_{\rm UV} $ and $\Lm_{\rm sc}  $ 
that specify the hierarchy and the initial conditions. 
Therefore, the emergent IR scale has to appear in dimensionless combinations with those scales.

We clearly see in Eqs.~(\ref{Solution_of_RegionII}) and (\ref{IRScale_MC}) that 
the initial energy scale $  \Lm_0 = \Lm_{\rm sc}$ in Region II 
results in the factor of $ \sqrt{eB} $ in the final result (\ref{DynamicalMass}), 
and also that the initial condition $ G(\Lm_{\rm sc}) $ results in the exponent. 
Therefore, the dependence of the mass gap on the magnetic field comes from 
that of the screening mass $\sim \sqrt{ \alpha eB} $. 
Also, we could obtain the logarithmic exponent in Eq.~(\ref{eq:QED_sc}) 
from the initial condition (\ref{solution_mgamma}) resulting from the evolution in Region I: 
The separation of the RG evolutions in the two regions 
naturally led us to take the initial condition $ G(\Lm_{\rm sc}) $ 
at the intermediate scale $ \Lm_{\rm sc} $ instead of at the UV scale $ \sim \sqrt{eB}$. 
Note that this exponent is independent of the magnetic field. 
This is because the only two available scales are both proportional to $ eB $.

The origin of the factor of $ \alpha^{1/3} $ is more subtle. 
In the passing from Eq.~(\ref{IRScale_MC}) to Eq.~(\ref{DynamicalMass}), 
we notice that a part of this factor comes from the initial energy scale $ \Lm_{\rm sc}$ in Region II 
and the other part comes from the initial condition $ G(\Lm_{\rm sc}) $ in the exponent. 
The latter contribution was correctly obtained as the result of the RG evolution in Region I, 
as discussed below Eq.~(\ref{solution_mgamma}). 
In a word, without the energy-scale dependence of the tree-level contribution in Eq.~(\ref{RGeq-I_MC}), 
one cannot reproduce this factor. 

The above observations clearly indicate the importance of 
respecting the interaction properties in the underlying theory. 
Therefore, it is useful to compare with the results in 
the NJL model (\ref{DynamicalMassNJL}) and the unscreened QED (\ref{eq:QED_unsc}), 
and identify the origins of the differences in the RG analyses.

{\it QED with unscreened photons}.--- 
Without the screening effect, the relevant RG equation is only Eq.~(\ref{RGeq-I_MC}) in the entire scale regions. 
The solution is given in Eq.~(\ref{GI_MC}). 
As the scale $\Lambda$ decreases, this solution hits the Landau pole. 
Therefore, the dynamical IR scale is obtained from the following equation: 
\beq
- \sqrt{ \frac{ \alpha }{ 2 \pi } } {\rm{log}} \frac{ \Lambda_{\rm{IR}}^{2} }{ 2eB } 
\simeq  \frac{ \pi }{2} 
\, ,
\eeq
which yields the mass gap 
\beq
m_{\rm{dyn}} 
\simeq 
\sqrt{2eB} \, {\rm{exp}} \left( - \frac{ \pi }{2} \sqrt{ \frac{ \pi }{ 2 \alpha } } \right)
\, .
\eeq
This result agrees with that from the SD equation (\ref{eq:QED_unsc}) up to an order-one constant factor. 
In unscreened QED, there is only one scale, that is, 
the initial scale $ \Lm_{\rm UV} $ for the entire RG evolution. 
This explains the overall factor of $ \sqrt{eB} $ and the independence of the exponent from $ eB $. 

More in detail, as we have already discussed, 
the screening properties are reflected in the exponent. 
Without the screening effect, the exponential suppression 
is parametrically weaker than that in the screened QED~(\ref{DynamicalMass}). 
The overall factor of $ \sqrt{eB} $ comes from the initial scale of the RG equation, 
of which the value is, however, different from that in screened QED: 
While it was $ \Lm_{\rm sc} \sim \sqrt{\alpha eB} $, 
it is now $ \Lm_{\rm UV} \sim \sqrt{eB }$. 
This also means the absence of the overall power factor of $ \alpha $ in unscreened QED.

{\it NJL model}.---
The coupling constant $ G_{\rm NJL} $ in the NJL model does not 
have any energy-scale dependence in its tree-level Lagrangian. 
Therefore, the RG equation for the NJL model is formally the same as Eq.~(\ref{RGeq-II_MC}) for Region II, 
but with an initial scale at the UV region $ \Lm_0 \sim \sqrt{eB} $. 
Since there is no photon propagator in Eq.~(\ref{effecG_magMC}), 
the effective coupling is, as the result of the Gaussian integral, given by $ \rho^{}_{\rm{LLL}} G_{\rm NJL} $ 
where $ \rho^{}_{\rm{LLL}} = eB/(2\pi) $ is the density of states in the LLL. 
Plugging the initial scale and initial condition into Eq.~(\ref{IRScale_MC}), 
one can precisely reproduce the mass gap (\ref{DynamicalMassNJL}).

Although this result is similar to that in QED, the magnetic field dependence is different. 
As discussed just above, the exponent in QED is independent of the magnetic field, 
because of the absence of the dimensionful quantities other than 
$ \Lm_{\rm UV} $ and $ \Lm_{\rm sc} $ which are both proportional to $ eB $. 
On the other hand, in NJL mode, the dimensionful coupling constant $ G_{\rm NJL} $ 
appears in combination with the factor of $ eB $ in the density of states $ \rho^{}_{\rm{LLL}} $. 
Therefore, the mass gap in the NJL model increases much faster than 
that in  QED with an increasing $ eB $ due to the diminishing exponential suppression. 
When discussing QCD, this behavior may not be regarded physical 
because of involved intrinsic energy-scale dependences 
in the intermediate- to low-energy QCD, 
which implies one of limitations of the NJL model. 

It will be interesting to investigate an extension to asymptotic-free theories 
which offer an additional intrinsic IR scale, i.e., the QCD scale $ \Lm_{\rm QCD} $. 
Indeed, there has been an important issue that, when $ eB \gtrsim    \Lm_{\rm QCD} $, 
typical effective models of QCD fail to explain the linear dependence of the chiral condensate on $  eB$, 
i.e., $ \langle \bar q q \rangle \sim eB  $ 
which was observed in the lattice QCD simulations~\cite{Bali:2012zg} 
(see also Ref.~\cite{Bali:2013cf} for a brief summary). 
It was pointed out that the dynamical mass should stay at $ m_{\rm dyn} \sim \Lm_{\rm QCD}  $ 
with increasing $B$, 
because the dimensional reduction leads to a factorization which is roughly $ \langle \bar q q \rangle \sim eB m_{\rm dyn} $~\cite{Kojo:2012js} (see also Refs.~\cite{Watson:2013ghq, Mueller:2014tea, Hattori:2015aki}). Also, to explain the inverse magnetic catalysis~\cite{Bali:2011qj, Bali:2012zg}, this saturating behavior of $m_{\rm dyn}$ is thought to be important for the thermal excitations not to acquire too huge a mass gap to restore the chiral symmetry~\cite{Kojo:2012js, Hattori:2015aki}. 
Thus, those issues at zero and finite temperatures appear to reduce to a problem of explaining the saturation of the mass gap with an increasing $B$. In Refs.~\cite{Kojo:2012js, Hattori:2015aki}, 
the IR-dominant interactions are shown to be 
important for reproducing the saturation on the basis of the SD equation. 
In the above RG analyses, it is clear that the mass gap has to be $ m_{\rm dyn} \sim \sqrt{eB} $ 
in any theory/model containing only $ \Lm_{\rm UV} $ and $ \Lm_{\rm sc}  $: 
For example, {NJL model may not work as an effective model of QCD 
in the study of the magnetic catalysis for this reason. 
It is then interesting to see how the intrinsic scale $ \Lm_{\rm QCD} $ modifies 
the chart of the hierarchy in Fig~\ref{fig:RG-MC} and 
manifests itself in the mass gap in the language of the RG 
and the functional RG~\cite{Skokov:2011ib, Scherer:2012nn, 
Fukushima:2012xw, Andersen:2014xxa,
Kamikado:2013pya, Braun:2014fua, Mueller:2015fka, Aoki:2015mqa}.

In summary, we have closely looked into the magnetic catalysis phenomena by means of the RG method 
with a special care of the scale separation in the RG evolution. 
Especially, we elaborated the treatment of the screening effect on the photon propagator, 
and  showed its crucial role in the determination of the dynamical mass gap. 
Our result on the mass gap agrees with that from the SD equation~\cite{Gusynin:1998zq, Gusynin:1999pq}. 


\vspace{0.2cm}

{\it Acknowledgement}.---S.O. thanks I. A. Shovkovy for useful discussions. 
K.I. and S.O. thank Y. Kuramoto for insightful discussion on related subjects. 
K.H. thanks Toru Kojo for discussions. 
K.H. and S.O. are grateful to KEK and Yukawa Institute 
for hospitalities and financial supports during their visits 
and  the YITP workshop ``Strangeness and charm in hadrons and dense matter,'' 
where a part of this work was achieved. 
The research of K.H. is supported by China Postdoctoral Science Foundation 
under Grant Nos.~2016M590312 and 2017T100266. 
The research of S.O. is supported by MEXT-Supported Program for the Strategic Foundation at Private Universities, ``Topological Science" under Grant No.~S1511006. 

\appendix

\section{Logarithm from the one-loop scattering amplitude}

\label{sec:amp}


Inserting the LLL propagator (\ref{eq:LLL-prop}) into 
the one-loop scattering amplitude (\ref{Amp_1-loop}), 
we have 
\begin{eqnarray}
&&\hspace{-1cm}
\M_1 = 
i (-iG)^2 \int \frac{d^2k_\para}{(2\pi)^2} 
\frac{ i [ \bar u (p_\para^{(3)}) \gam_\para^\mu \sla k_\para\prj_+ \gam_\para^\nu 
u (p_\para^{(1)}) ]} { k_\para^2 + i \ep}
\nn
\\
&& \hspace{1.2cm} \times
\frac{i[v (p_\para^{(4)}) \gam_{\para\mu} ( \sla k_\para - \sla P_\para) \prj_+
\gam_{\para\nu} \bar v (p_\para^{(2)})]} {  (  k_\para - P_\para)^2 + i \ep}
\, .
\nn\\
\end{eqnarray}
To proceed, we first perform the contour integral for the $ k^0 $, 
which is given by the contributions from the residues of four poles at 
$  k^0 = \pm k_z - i \, \sgn(k^0)  \ep $ and $ k^0 = \pm k_z + P^0 \mp P_z - i \, \sgn(k^0)  \ep $. 
After inserting these poles into the numerator, 
one finds that the chirality projection operator $ \Q_\pm = (1\pm  \gam^5)/2 $ 
naturally arises through an identity 
\begin{eqnarray}
(\gam^0 \pm \gam^3) \prj_+ = (\gam^0 \pm \gam^3) \Q_\pm
\, .
\end{eqnarray}
To pick up the contributions to the scatterings between 
the particle and antiparticle pairs carrying opposite chiralities, 
one can use useful formulas 
\begin{eqnarray}
\gam_\para^\mu \gam_\para^\alpha \gam_\para^\nu 
= g^{\mu\alpha}_\para \gam_\para^\nu - g^{\mu\nu}_\para \gam^\alpha_\para 
+ g^{\alpha \nu}_\para \gam^\mu_\para
\, ,
\end{eqnarray}
and 
\begin{eqnarray}
 (\gam^0 \mp \gam^3) \Q_\pm =  (\gam^0 \mp \gam^3) \prj_-
 \, ,
\end{eqnarray}
where the LLL spinors are orthogonal to $ \prj_- $, i.e., $  \prj_- u= \prj_- v=0 $. 

After performing straightforward algebraic calculation, 
the scattering amplitude is obtained as 
\begin{eqnarray}
&& \hspace{-0.9cm}
\M_1 =  
G^2 
[\gam^\mu_\para]_{\rm R} [\gam_{\para\mu} ]_{\rm L}
\int \frac{dk_z}{2\pi}  
\frac{ \theta(k_z-P_z) - \theta(- k_z)  }{ k_z - (P^0  + P_z)/2 }
\nn\\
&&
+ G^2 [\gam^\mu_\para]_{\rm L} [\gam_{\para\mu} ]_{\rm R}
\int \frac{dk_z}{2\pi} 
\frac{  \theta(k_z) - \theta ( - k_z + P_z ) } { k_z + (P^0 - P_z)/2}
,
\nn\\
\end{eqnarray}
where the shorthand notations $ [\gam^\mu_\para]_{\rm R/L} [\gam_{\para\mu} ]_{\rm L/R} $ 
denote 
$[ \, \bar{u}_{\rm{R/L}} (p^{(3)}_{\parallel})  \gamma_{\parallel \mu}
u_{\rm{R/L}} (p^{(1)}_{\parallel}) \, ]
 [ \, v_{\rm{L/R}} (p^{(4)}_{\parallel} ) 
 \gamma^{\mu}_{\parallel} \bar{v}_{\rm{L/R}} (p^{(2)}_{ \parallel}) \, ]  
 $, respectively. 
The LLL spinors with the definite chiralities are 
$ u_{R/L} = \Q_\pm u $ and $ v_{R/L} = \Q_\pm v $. 
Since the energies and momenta of the scattering particles are less than $ \Lambda  $, 
we have $ P^0,P_z \lesssim \Lambda$, and thus the logarithmic contribution 
arises in the form of the integral (\ref{eq:kz}) for the both scattering channels 
up to a numerical factor. 

\bibliographystyle{apsrev4-1_r}
\bibliography{bib}

\end{document}